# Hidden Underbelly of the Silicon Valley: Algorithmic Exploitation and Health in Data Work Value Chains

**Mophat Okinyi**[1], **Richard Mathenge**[1], **Mohammad Amir Anwar**[2,3]

[1] *Techworker Community Africa, Nairobi*
[2] *University of Edinburgh*
[3] *University of Johannesburg*
*Email: mohammad.anwar@ed.ac.uk*

## Abstract

With robots expected to replace humans in some professions, AI presents a new development prospect through the provisions of data work. For over a decade, large Silicon Valley technology firms have been relying on outsourcing of data work via a host of intermediary suppliers and labour platforms to different parts of the globe often the Global South region, such as East Africa. Much of this work is often shrouded in secrecy as firms rarely reveal the extent of their value chains. This results in the poor and marginalised forming the hidden underbelly of the Silicon Valley, training some of their most advanced machines in adverse working conditions. Drawing upon the survey of workers in Kenya, a major hub for data work in Africa, the paper highlights the physical and psychological impacts on workers. Survey data is complemented with in-depth interviews and auto-ethnographic account of two ex-data workers-turned activists who worked for a large data enrichment firms in Kenya. Overall, the study highlights serious mental and physical health issues experienced humans behind the making of AI systems.

## Introduction

> *Back in 2020 during COVID-19, I got laid off from work. It was the worst period to experience unemployment. A former colleague reached out to me about a vacancy that opened within the organisation they worked at. It was Sama. Since I had nothing to lose, I put in my application and was invited for an interview. It was at this point that I found myself in the murky world of data work (Richard).*

> *The room where we did content moderation became my world for hours every day. It had few windows and lacked fresh air, with cold lighting and rows of desks with computers that never stopped feeding us tasks. The workplace felt more a prison cell than a workplace. I was exposed to child sexual abuse, violence, and hate speech daily. The management told us to "treat it like data", as if this could shield me from emotional harm. But no amount of professional detachment could protect my mental health from the sheer volume of horror I was reading and viewing (Mophat).*

For machines dependent on artificial intelligence (AI) to work efficiently, they need to be fed massive amounts of training data. This is called machine learning (the science of getting computers to make decisions) whereby data enrichment tasks (e.g., text prediction, image and video annotation, speech to text validation, and content moderation, also commonly known as data work) is carried out by behind-the-scenes human labour (see Miceli and Posada, 2022). A large body of work has emerged exploring the ways in which data work is enabling workers in some of the poorer parts of the world to get access to livelihood opportunities (e.g. Tubaro et al., 2025; Soares Seto, 2025; Anwar, 2024a; Gray and Suri, 2019). Large Silicon Valley firms rely on outsourcing of data work to different parts of the planet via a host of suppliers (e.g., Scale, Sama, Remotasks, Stepwise, Landing.AI, Factored.AI) and digital labour platforms (e.g., Appen Amazon Mechanical Turk, and Upwork) to train their AI systems. These suppliers and labour platforms tap into workforce often in some of the most marginalised regions of the world (Tubaro et al.., 2025; Anwar 2024). This is how data workers like Mophat and Richard in East Africa get plugged into the AI value chains.

Participation in AI value chains is seen by many e.g. policy makers, private corporations and international organisations alike, as as a route towards economic development, widely referred as AI for development (AI4D) (see Toupin and Siad, 2025). In the context of outsourced data work, it is seen having the potential to lift millions out of poverty in regions like Africa with rampant unemployment (Nicholson et al., 2018). Yet, scholarship on data work has raised serious alarms over its economic development potential (e.g. poverty reduction via provision of wages) (e.g. Gray and Suri, 2019; Anwar and Graham, 2022; Muldoon et al., 2023; Le Ludec et al., 2023; Altenried, 2020). A central concern in this scholarship is the ways in which data work value chains are structured and organised and how it affects workers (Anwar, 2024b).

Big tech firms in the global production networks or global value chains have always engaged in outsourcing and offshoring practices due to the competition in the market (Dicken, 2015). These production practices adversely impact working conditions in the value chains. Ben Selwyn (2019) calls them poverty chains, whereby workers get integrated into value chains and yet forced to work under exploitative conditions including depressed wages, workplace injuries, and abuse by the management. Data work also gets outsourced via complex webs of actors spread across global value chains. The underlying argument we are making here is that conceptualising data work as an extension of the wider outsourced value chains of digital services work (i.e. a variety of outsourced information technology services including back-end admin work, customer support, sales, web development, transcription, content generation, search engine optimisation, etc.) allows us to understand better the ways it affects workers. Where the paper departs from the existing literature on data work is its focus on non-economic (i.e. health) impacts on workers in Africa.

Outsourced IT services value chains, primarily built on labour arbitrage, are notorious for adverse working conditions (Peck, 2017). By way of example, call centre work is highly intensive and performed under extensive workplace monitoring (see Taylor and Bain, 1999). Data work is also performed under extreme workplace surveillance, with some referring it as ‘digital Taylorism’ (Anwar and Graham, 2022). Data work can be generally mundane and highly repetitive coupled with strict demands put by management means work intensity is high.

Essentially, data workers act like janitors to do the dirty work (Irani, 2019) while suffering from physical and mental health problems. There are reports of over a hundred workers diagnosed with severe PTSD, according to a 2022 lawsuit filed in by leg Motaung, a data worker in Kenya, against Meta and Sama (the company doing outsourced content moderation work for Meta in Kenya). In early 2025, a Nigerian data worker committed suicide in Kenya, where she was working for a large outsourcing company called Teleperformance. Data workers themselves have been reporting about their work's toll on mental health in the last couple of years (e.g. Malgwi, 2025; Gebrekidan, 2024). This paper builds on some of this emerging evidence to examine the scale and scope of physical and psychological health impacts of data work in East Africa, an aspect less studied so far. The central argument it makes is that AI is built through the exploitation of both bodies and minds of data workers. The argument is supported through survey evidence among Kenyan data workers along with ethnographic work with data workers in Nairobi.

The remainder of the paper is structured as follows. The second section of the paper develops a framework to understand how and why data work influences working conditions and what it means for health and well-being of data workers. It draws on literature around value chains of outsourced digital work (e.g. in platform economy and business process outsourcing industry) to explain health impacts at work. The third section discusses methodology by grounding it in a well-acknowledged practice of workers inquiry (Marx, 1880; also, Haider and Mohandesi, 2013; Woodcock, 2014). In the fourth section, we elaborate on the physical and socio-psychological impacts faced by workers evidenced through survey and interview data. This section also integrates first-hand account of the data work industry through Mophat and Richard's central role on the frontline of AI development. The conclusions tease out the implications on what it means to resist AI and underscores the need for further extensive research on health-related issues in data work value chains.

## AI-Labour Nexus[1]

Large tech firms (i.e. lead firms) from Microsoft to Amazon to Alphabet to Meta, are developing large language models (LLMs) via training of large corpus of data from various digital repositories on the internet. Some of their LLMs (e.g. ChatGPT and Gemini) have been remarkably successful in showing high-performance such as passing the medical and law exams. But this success is dependent on the human feedback loop, i.e. training via supervised finetuning (SFT) and Reinforcement Learning with Human Feedback (RLHF) techniques (see Liu et al. 2023; Steinnon et al., 2020). These activities are outsourced and subcontracted to a wide array of actors, including data training firms and labour platforms (Anwar 2024; Muldoon et al., 2024).

Data training firms are best understood as intermediaries between the lead firms and workers performing a wide variety of data enrichment tasks (verifying, annotating, labelling, imitating, etc. see Tubaro et al, 2020). Microsoft has a partnership with OpenAI, the company responsible for developing the ChatGPT (Duhigg, 2023). OpenAI outsourced some of the training for its large language models such as ChatGPT to Sama, a US-based training firm. Sama's sales team is based in the US where they secure contracts with large tech firms. The work is then sent across to their delivery centres in some of the poorer parts of the world where thousands of data workers do the training work (i.e. annotation, curation, classification, and verification of data). Two of the main ones are in Kenya and Uganda, and previously had centres in Haiti and Ghana as well. Similarly, almost all the major social media firms (i.e. Meta, Twitter, Tik Tok) requires thousands of workers to keep their websites clean from pornography, violence, child abuse, etc. These workers are called content moderators, who are hired by supplier firms (e.g. Accenture and Covalen) from Bangalore to Nairobi to Bogota. Essentially, these specialist firms are part of the global value chains of the outsourced IT services sector.

[1] This section builds on author's previous work on data work value chains (Anwar, 2024b).

The global outsourced IT services sector employs an estimated 50 million workers, according to a 2025 report by Caribou Digital and Genesis Analytics. On the African continent, there are about 1 million workers in this sector (Caribou Digital and Genesis Analytics, 2025). In Kenya there are just over 34 thousand employed in the sector, according to the 2024 Africa Global Business Benchmarking and Market Report Data Bank. These include call centre work, technical support, software development, data management, sales, accounting and data work. The report does not disaggregate data based on different types of work. However, it does note in Kenya just over 23,249 thousand workers are serving the domestic market and over 11,597 are serving the international markets. While the demand for much of the data work in Kenya comes from abroad, it is hard to put an accurate figure on how many workers are employed by firms (e.g. Sama, Cloud Factory, Teleperformance, etc.) doing data work in the country.

Apart from these data training firms there are other kind of players involved in outsourced data work value chains: labour platforms. These include Upwork, Amazon Mechanical Turk, Appen, Scale AI, Remotasks, etc., which allow businesses and start-ups, researchers to hire on-demand workers globally to do data work for them. While much of the demand originates in the global North (or the richer countries) the supply of workers on these platforms is concentrated largely in the global South regions (or poorer parts of the world), including Africa (see Kässi et al., 2021; Anwar et al., 2024).

According to the World Bank, there are an estimated around 154 million unique registered platform workers, of which 60 million are active. This number, according to World Bank can be as high as 435 million (Datta et al., 2023). The Online Labour Observatory data suggests that based on on 162 of the 351 labour platforms there are 163 million registered user accounts or workers in the world (Kässi et al., 2021). Not only these estimates have limitations, these include a variety of work on platforms. Hence, it is difficult to say with clarity how many data workers are there in the world or for that matter across the whole of Africa or just in Kenya.

Data training firms and labour platforms can be thought of as the equivalent of modern-day business process outsourcing (BPO)

firms with their production activities spread out around the world. Miceli and Posada's (2022) research on data work in Latin America highlights the presence of BPO firms doing data generation projects for the US clients. Anwar's (2024b) study on training centres in Africa reveal several foreign data training firms operating in Africa label themselves as BPOs to get data work contracts from large Silicon Valley tech firms. There is also presence of local Kenyan BPO firms doing subcontracted data work through foreign firms such as Sama (Anwar and Graham, 2022). The argument we are making here is that data work can be thought as a newer extension of the outsourced services value chains. Data training firms and platforms are the new actors in these value chains. This kind of conceptualisation allows us (a) to understand the value chains or production networks of AI better, particularly how data work gets done and in which places and (b) to better explain the exploitative experiences of data workers as described by Mophat in the opening paragraph of the article. Workers not only get exposed to extremely violent and abusive material but also feel trapped into the vicious cycles of exploitation.

Much of the literature to date on data work in Africa has studied economic impacts such as the wages, pension and medical insurance, etc (e.g. Muldoon et al., 2023; Posada, 2022), along with concerns around the labour process including algorithmic surveillance of work. Theft of wages are common (Bird and Schepers, 2025). Workers rarely have avenues to challenge their employers or have freedom of association. Within this literature, there is a consensus emerging that working conditions in this industry are generally poor, despite the early corporate narratives around impact sourcing and the endorsement of international organisations such as the World Bank (e.g. World Bank, 2016; Rockefeller Foundation, 2013). Sama, one of the key players in data work value chains, and headquarters in California, prides itself building 'ethical AI supply chains' (Gaddoniex, n.d). It claims to pay above national minimum wage and provide pensions and health insurance to employees in its training centres in Kenya. A recent study on the 'impact sourcing' initiatives of firms like Sama found that purported benefits to workers remain marginal (Muldoon et al., 2023). A 2024 investigative report by the Time magazine found working conditions to be worse than previously thought, including Kenyan workers getting paid less than US$2 per hour. In Madagascar, data work is found to be associated with low income and long working hours (Le Ludec et al., 2023). The paper, however, explores another aspect that has received less scholarly interest so far, i.e. physical and mental health impacts.

### *Physical and mental health at work*

For millennia, people have performed a variety of tasks to fulfil their material, social, psychological, and physical needs (Jahoda, 1984). Most important of them is paid work's connection to wages or monetary compensation that allows workers to fulfil their immediate material needs such as food, and shelter. These needs are more important if the workers are from socio-economically poor backgrounds. Over half of the working population on the African continent is considered working poor, i.e. earning less than $3.65 per day (ILO, 2025). For African workers then, earning a livelihood to put food on their table is of utmost important. But beyond the economic

imperative, work affects the health of workers too.

Scholars have maintained that employment protects and fosters health (e.g. Ross and Mirowski, 1995). Social interactions between fellow workers, family, and friends have long been considered effective for the psychological and physiological health of workers (Jahoda 1982). However, precarious employment arrangements (i.e. informal work or shift work; high job insecurity; lack of autonomy) can lead to lower psychological and physical well-being of workers (Law et al., 2020; Cheng et al. 2008). For example, shift work is associated with insomnia, substance abuse and even suicidal thoughts (Brown et al., 2020). Long working hours negatively affect sleep quality, affecting workers and firms' productivity (Afonso et al., 2017). Prolonged exposure to stressful work can lead to psychological distress, including anxiety and burnout (Dobson and Schnall, 2009). As we demonstrate in the remaining of this paper, data work value chains are characterised by precarious working conditions, which can lead to serious health issues.

## Methodology

The paper adopts a mixed methods approach combining surveys with in-depth interviews and the auto-ethnographic account of two ex-data workers in Kenya. Secretive nature of the data work value chains means accessing workers remains a challenge. Furthermore, workers fear for their livelihoods, hence trust remains a major issue for researchers. Leveraging social media and networking sites can allow researchers to find respondents (Liu et al., 2016). However, these strategies have self-selection bias compromising its quality (Lehdonvirta et al., 2021).

To overcome these problems, we adopted two-pronged strategy. First, we collaborated and partnered with data workers turned activists in Nairobi. These data workers helped develop the survey questions and provided feedback on them. A short survey was designed on Qualtrics to be completed in under 10 minutes. Data workers distributed surveys into their community WhatsApp group in Nairobi in February 2025. This allowed targeted surveys of data workers. Secondly, no compensation was offered to workers to complete the survey. This was a deliberate move to avoid respondents who might be driven by monetary rewards, a concern known among scholars to generate poor quality data (Wolf and Anwar, 2025).

A total of 229 responses were received and 124 of those responses were usable. Therefore, while the results of the survey are not representative of data workers in Kenya, the sample provides extensive insights into hard to reach and invisible human labour behind AI. Kenya is one of the largest markets for the data work in Africa. Several leading data training firms are present here, e.g. Sama, Digital Divide Data, Cloud Factory, Teleperformance, Stepwise, and Remotask.[2] Kenya is also one of the largest African suppliers of labour on platforms such as Upwork, Fiverr, and Appen for data work. Therefore, the sample and findings are indicative of the wider trends among African data workers in the industry and provide an entry point to further explore the experiences of human labour in AI value chains.

[2] Remotasks shut their Nairobi-based operations in 2024)

We then conducted in-depth interviews with 10 survey respondents in Nairobi in June 2025. Only those who agreed to be interviewed were contacted for interviews. We further selected respondents for interview to ensure diversity including gender, types of work, length of employment, and variety of health impacts. Interviews allowed us to raise new questions and seek clarifications on several themes emerging from survey responses. Interviews lasted between 55-75 minutes. All interviews were audio recorded and transcribed for coding on NVivo for themes such as work intensity, working hours, health impacts, access to medical facilities, welfare provisions at work, etc. Names of the respondents have been anonymised.

The paper further adopts an auto-ethnography approach with two of the co-authors as ex-data workers. This kind of approach has been widely utilised by researchers to investigate their workplaces (e.g. Cant, 2019; Woodcock, 2016). The approach in the paper can also be understood as a form of workers' inquiry (Marx, 1880), where workers lead the knowledge production with a view to understand the inner workings of capitalism and organising against it (see Notes from Below, 2018). The two co-authors bring inside information to the ways in which data work value chains get organised, the ways in which workers get integrated in them, and the personal experience of data workers including impact on health. Richard and Mophat's stories also show how poor working conditions in the data work value chains can ferment seeds of resistance. In the emerging literature on data work, there is a recognition of the value of these kinds of methodological initiatives which enable a better understanding of the ways labour is exploited at the hands of capital and how solidarity emerge. The paper, therefore, sits in close dialogue to the Data Workers' Inquiry project (at the Distributed Artificial Intelligence Research Institute (DAIR)) where workers are reporting their own working conditions and naming and shaming unscrupulous employers and platforms (See Miceli et al., n.d).

Finally, gender has a strong bearing on the experiences at work, and we are mindful of these dynamics. For example, in the world of platform economy, it has been widely reported to affect women adversely (e.g. James, 2025). Our approach in this paper does not specifically discuss findings along the gender dimension as this is a subject of our ongoing collaboration with female data workers for a future paper.

**Who are Kenyan data workers?**

In our sample, a majority of the respondents were male (62.2%) compared to females (36.8) and 1 person indicated to be non-binary. Gender divide remains a prominent issue with women underrepresented in the digital economy activities (Gillwald and Partridge, 2022). Furthermore, over 92% of our respondents were between 18-34 years of age, indicating data work's ability to attract younger workforce (*Figure 1*). The digital economy activities tend to be associated with younger generation. A 2021 survey by the ILO found that most of the workers on digital labour platforms are primarily between the ages of 18-35. Similarly, a European survey found more young people between the ages of 15 and 29 are likely to be working on digital platforms than people aged 30-64 (Piasna et al., 2022). Beyond the platform economy, Dhanpat et al., (2018) found half of the employees in one of the largest call centres in South Africa were between 18-24 years old.

Additionally, a large share of our respondents had diploma (48.3%) and secondary education (12.2%). While the rest of the respondents had some sort of college or University degrees, with 34.4% with undergraduate degrees and over 4.9% with post-graduate degrees. Two of the co-authors of the paper have university degrees. Richard came from public relations background and Mophat has a degree in education. While Richard had worked in varied sectors including insurance and oil, before moving into customer services working for Technobrain[3], Mophat's first job was in the data work industry with Sama in Nairobi.

Over 42% of our respondents were formally employed with a majority working for Sama, Teleperformance, Cloud Factory (all international firms). A third (34%) of respondents were doing data work via labour platforms such as Appen, Remotask, and Upwork. A sizeable share of data workers was combining employment at a firm and freelancing on platforms. This is a well-known practice known as hustling. While hustling is understood as a combination of two or more jobs or activities for daily survival (see Ravenelle, 2019), in urban Kenya hustling is understood as both struggle and agency (see Thiem, 2021).

Firms, looking to tap into this aspirational but cheap workforce, also sell these workers the idea of doing social good, i.e. helping build advanced technologies of tomorrow. Workers buy into this narrative. For young African workers digital jobs are tied to the imaginaries of being part of the 'global village' and a sense of connection the wider world (see Anwar and Graham, 2022; Burrell, 2012). A data worker told us she joined Sama because this work was considered to have good reputation locally and seen by many as a path to economic success. Instead, she was doing data labelling from a discreet suburb in Nairobi, clicking repeatedly on thousands of images daily to train algorithms (Interview Nairobi June 2025). In Brazil, Grohmann et al., (2022) aptly describe this kind of work as 'click farms'. In the African contexts, these click farms take the form of old shipping containers (Anwar and Graham, 2020). Except old containers are not being used for trade in material products like mineral ore, shoes, garments, toys or foodgrains, but the extraction of economic value through the exploitation of labour force.

***Algorithmic exploitation and health***

One of the defining features of the new digital economy jobs is the widespread use of advanced socio-technological infrastructures by the management or employer to control and monitor workers and labour process (see Jarrahi et al., 2021; Schor et al., 2020). It is referred as algorithmic management (whereby employers deploy digital surveillance and monitoring software to control workers and the labour process). The term is often used to explain how and why workers experience poor working conditions (Stark and Broeck, 2024). What we are arguing is that data workers experience *algorithmic exploitation*. Data workers not only train and develop the most advanced algorithms for AI, but they are also controlled and exploited by the very algorithmic system they help built. We are not claiming that it is the algorithm stupid. Instead, the management or employers are exploitative. Employers carefully design workplace management systems to maximise the extraction of economic value from labour. For example, deliberate attempt to set unrealistic targets which keep workers at their work desk for long hours, penalising for time spent on non-work activities, etc. This kind of set up has precedence in outsourced services sector, especially in call centre industry (see Woodcock, 2016; Taylor and Bain, 1999). As we noted above the presence of BPOs in the data work value chains suggest some organisational practices and workplace cultures get exported to data work. Essentially algorithms get trained on the back of worker exploitation. Below we outline how algorithmic exploitation gets embedded in data work value chains. One of the central aspects of data work is the high intensity of working, which is the source of algorithmic exploitation.

[3] Technobrain is a large regional information technology services firm with presence in multiple countries within Africa and the Gulf region.

Data workers in Kenya are primarily involved in data annotation and content moderation work (*Figure 2*). Data annotation involves workers labelling images and videos correctly based on specific guidance and instructions given by the management. This work can be done in seconds or in some cases take several hours for annotating a single image (Anwar and Graham, 2020). Content moderation involves workers reporting and tagging harmful content on the internet including social media channels such as Twitter, Facebook, and Tik Tok. This work is done in real time, and workers have to be in-front of the computer screens to receive new social media posts and are usually given a few seconds to flag it for removal or approval. Content generation and transcription are two other key work activities performed by Kenyan workers. Kenya has long been the hub of transcription activities, particularly via major labour platforms (Kassi and Lehdonvirta, 2018) or via specialised transcription companies, most of which feeds into automatic transcription products of large companies including Netflix and YouTube.

*Figure 3* shows common issues faced by data workers. Two concerns reported by most respondents were delayed payments and high workloads. Both content moderation and data annotation work can be highly intensive with unrealistic demands and targets imposed by the employer (Chandhiramowli, 2024; Miceli and Posada, 2022). This results in long working hours and overtime (without additional pay). As Mophat confirmed:

> *The targets were very crazy. We were told to hit a target of 1000 images in a day, which can be difficult to achieve. The managers were harsh, and they would not listen to workers. If we complain about targets, they would threaten to send us home. We had no options. As a result, many would be in the office past midnight…If you want to complete 1000 data sets, you will have to do over time. Instead of working for eight hours, you'll work for 10 hours. You will not get enough time for breaks. We had strained posture and get regular back pains. Our legs would get numb from sitting for long hours.*

Several interviewees told us that they would lose a portion of their wages if they failed to meet daily targets. This kind of highly intensive work environment is not unique to data work. In fact, high intensity working conditions have been extensively documented both in the call centre industry and the platform economy (e.g. Woodcock, 2016; Taylor and Bain, 1999; Wood et al., 2019; Anwar et al., 2024). What makes the matter worse for data workers is the exposure to harmful content at work.[4]

Content moderators (2nd largest share of respondents) are exposed to traumatic content on a regular basis. However, it is surprising to see fewer respondents in our sample highlighting exposure to traumatic content as their main concern. One reason could be workers' subjective preference. In the context, such as Kenya, where poverty is an everyday reality for many, data workers immediate priorities were economic incentives at the cost of their health. In the words of a female data worker:

> *I was more concerned about paying my rent. Bringing food to my young family was more important to me. Hence, I would try to block all the filth I was getting exposed to at work. I was hoping it will fade away with time. But it did not* (Interview Nairobi, June 2025).

Another data worker said, *we were willing to sacrifice our mental health over livelihoods. It's a mindset here* [in Kenya] *that you have no option* (Interview Nairobi, June 2025). For over 44% of our respondents the main motivation to do data work was better wages and another 32% said there are lack of decent work opportunities in their local labour markets. The argument here is that workers' subjective preferences may influence how they relate to serious concerns around their work, for example health.

Nonetheless, data work's impact on workers' health remains a serious issue. We asked the respondents to indicate what kind of health-related issues they have faced at work. *Figure 4* shows most common being exhaustion (37.7%) and mental stress (30.3%). This is followed by insomnia (14%), vertigo (10.8) and weak eyesights (3.2%) In terms of frequency of stress, over 32% reported to have experienced stress often, while above 35% experience it sometimes, 30% said they rarely experience stress, and just 4% reported to have never experienced stress.

The long-term impacts of stress on workers' physical and mental health are equally important. Use of substance abuse, including alcohol and drugs (not including tobacco and caffeine) because of workplace practices has been widely reported in literature (e.g. Frone 2008; Harris, 2004). Factors such as increased workloads, night shifts, and fatigue are known to influence substance abuse among workers to overcome stress (see Cousin 2022; Bell and Hadjiefthyvoulou, 2022; ILO, 2003). In a high intensity work environment such as the outsourcing industry, the risk towards the increased use of other hard drugs and alcohol becomes significantly higher. During the interviews, we were told that there are numerous cases of workers developing addiction to drugs and alcohol. One interviewee told us that *'after our shift, some of us would go to a nearby pub to hangout. This was our way of sharing and discussing what we do at work but also lighten our mood. Little did we know that this habit would made many of us drinkers.* Another interviewee said, *'alcohol and drugs were used as coping mechanisms by*

[4] Call centre workers do experience customer cyberbullying (see D'cruz and Noronha, 2014).

*workers. It became such a big problem that workers were reporting at work drunk'* (Interview Nairobi, June 2025). Turning up to work while hungover has been reported in other sectors too including manufacturing and professional services (Walker and Bridgman, 2013; Mangione, et al., 1999; Ames et al., 1997). Walker and Bridgman (2013) noted drinking often means to overcome anxieties induced due to work or bonding with work colleagues. In data work, drinking became a way to overcome the trauma workers are exposed to*. We were 'trauma bonding',* as one ex-data worker put it with laughter (Interview Nairobi, June 2025)*.* This laughter could be understood just as dark humour or a way to suppress the deep emotional toll of their work.

An interviewee, who is not in the data work industry anymore, asked us to stop recording and ended the interview. For her, the refusal to share details about work experiences was not about confidentiality or privacy issues, but because discussing work was traumatising for her. Both Mophat and Richard also shared similar sentiments of lingering trauma they experience daily. Richard told us,

> *'We were watching content about bestiality, where intercourse is happing between animals and humans. We were expected read and watch this content. Material on child sexual abuse was very common. Children in my neighbourhoods are friendly they would run to embrace me. I remember one time coming home from work and this girl just ran up to my arms and all I am thinking is the stuff that I have just watched for the whole day. In my head, I am questioning what if it was this child in my arms. It was very disturbing.'* (Interview Nairobi, June 2025)

Mophat further added,

> *There were contents about rape and sodomy. We witnessed necrophilia. A single mother with a young boy and she is dead in the hospital. There were people having sexual intercourse with that woman. Developers and big firms did want the public to see, and they are trying to pull them down as much as they could. And we were the ones hired to clean it up for them.*
>
> *But they did not warn me about the effects of that work. We were not trained to deal with the graphic content. We were reading and viewing those 10 hours a day for months. These images stick to your mind. I had a young daughter born at the time. I will come back from work and reflect on what I have seen and read about people molesting kids. Then I would see my own child and all those things would start to go through my mind and I started to avoid my own daughter. I would go to sleep and start seeing those things again in my head. I lost my sleep had nightmares. I was dreaming about people sodomising me and would wake and scream and struggled to go back to sleep. I was not the only one there.'* (Interview Nairobi, June 2025)

Both were later diagnosed with PTSD in 2022. According to Richard, *'PTSD is quite common in the industry. But not everyone gets diagnosed in Kenya because very few would go to see doctors or can afford to see a specialist.* According to Afrobarometer survey, six in 10 Kenyans do not have medical insurance with majority reporting they cannot afford it (Afrobarometer, 2025). Over 66% data workers reported to have no access to mental health support or counselling, compared to 34 % saying yes. Richard confirmed that he '*reported health issues to the senior management about the conditions in which workers were doing the work. Sama agreed for psychological sessions or wellness sessions.*' However, at Sama there was only one counsellor for several hundred workers, meaning the waiting list was long and workers simply ignored it. For freelance data workers on platforms, the luxuries of medical insurance or other employer welfare benefits (e.g. holiday pay) rarely exist. So, what does participation in data work value chains means for Africa and African workers.

First, governments around the continent are increasingly ambitious towards outsourced IT services sector for job creation. In the Kenyan context, where informality is an everyday reality (ILO, 2018), the government's focus on outsourced services sector (e.g. data work) could be seen an attempt to expand formal sector employment in the country.[5] But the boundaries between formal and informal sector are not that distinct (see Meagher, 2019). Formal sector jobs also have informal work arrangements. For example, wages in call centres are extremely low, hiring and firing of workers are quite common, and workers perform their tasks in extremely controlled and monitored workplace (Anwar and Graham, 2019). Some segments of the data work value chains (e.g. data training firms) may appear to be formal (as far as the existence of a written employment contract). Others such as those doing work on labour platforms have limited opportunities for written contracts. For firm-based workers, their work is recognised as formal and yet they do not enjoy the benefits that come with formal jobs (job security, employee benefits, collective bargaining, etc). This creates a double whammy for data workers in Africa. The need for paid work keeps them in the sector despite the exploitative conditions in data work value chains. It also raises questions about the national economic development agendas favoured by governments which primarily focusses on the quantity of employment and not necessarily the quality of work.

Secondly, the outsourcing sector landscape is changing in the country (see Kleibert and Mann, 2020; Anwar and Graham, 2022). Local firms have been acquired. For example, Daproim, a local outsourcing firm (which did mainly sub-contracted work for digitization and data labelling for Sama) was acquired by a large US firm called Stepwise in 2020. Simbatech, a local outsourcing firm was acquired by CCI Global in 2016. Some international firms have emerged. Marjorel's Kenyan operations were acquired by a French outsourcing giant Teleperformance in 2023, which is now expanding its capacity, giving much needed boost to job creation locally. On the one hand, these trends point towards a growing international services sector in the country and hopefully more jobs. As one recent study by Caribou and Genesis Analytics (2025) noted the outsourced IT services sector in Africa has the potential to grow up to 1.8 million jobs by 2030. On the other hand, a highly sub-contracted value chains means workers increasingly find themselves embedded in segmented labour markets (Peck, 1996) and precarious employment relationships.

[5] According to the Kenyan National Bureau of Statistics, the informal sector has grown in the last five years faster in comparison to formal sector. Over 83% of the workforce in Kenya exist in the informal sector equating to 17.1 million people. This rate is lower than some of its neighbours such as Uganda (WIEGO, 2025) but it is more several times more than South Africa (Statistics South Africa, 2025). Informal sector jobs are characterised by low pay, wage theft, high levels of insecurities (both job and income) in general.

Thirdly, work in these sub-contracted value chains is by design extensively monitored and controlled by management (see Zhang et al., 2025). While this kind of workplace control is not new, advanced technologies have allowed the management to exert far more extensive control on workers and labour process than seen either in the assembly line production of the Fordist era (e.g. Fairris, 2002) or the assembly line in the head of the post-Fordist era (i.e. in the call centre industry) (Taylor and Bain, 1999). Even though data workers exist in some of the worst working conditions imaginable, this does not foreclose the possibility of them pushing back against their employers.

**Resisting AI from Africa**

In industrial relations theory, workplace exploitation and injustices are considered to generate antagonism among workers towards employers which lay the foundation for resistance in various forms (Edwards and Hodder, 2022; Ackroyd and Thompson, 2022). In the absence of meaningful initiatives by either the employers or the government to address poor conditions, several brave data workers have taken upon themselves to show that resistance against AI is possible.

One of the critical tools at hand is the refusal to work. This has precedent in Africa. Young people in Johannesburg have actively demonstrated their willingness to refuse to work for low-paid jobs (see Dawson, 2025). Similarly, in Egypt, young particularly from the educated backgrounds to stay unemployed and wait for longer periods in the hopes of better livelihood opportunities to come through, often in public sector (Assaad and Ehab, forthcoming). In both cases, these are deliberate strategies among African workers to refuse underpaid or poor-quality work. For African data workers refusal to work could also mean leaving the unscrupulous employer and joining a new one most likely in the same industry. This is referred as job hopping, which is quite common in the outsourced services value chains (Vira and James, 2012). A practice also confirmed by most interviewees who have changed their employers and are working for a new one. Despite a growing sector, data work in Kenya is still considerably small with a handful of employers and platforms operating. For example, workers either leave Sama and work for Cloud Factory or Teleperformance or vice versa. Data workers employed by these firms are also likely to have freelancing accounts on platforms such as Appen or Upwork. They have also turned to activism to leverage various sources of power, including symbolic power (see Webster and Dor, 2023).

African data workers (including two co-authors of this paper) have spoken about their experience and their testimonials are available in the public domain, including through Data Workers Enquiry project (Miceli et al., n.d.). They have launched lawsuits against big tech and data training firms like Sama. In 2022, Foxglove, a UK-based non-profit firm working with a Nairobi based team of lawyers and ex-data workers at Sama launched a court case against wrongful dismissals and poor mental health for workers. By 2024 the Kenyan Court of Appeal had ruled that the case can proceed to trial and that Facebook and Sama both can be sued in Kenya (Foxglove, 2024). Both Richard and Mophat, along with dozens of other data workers and activists, including Kauna Malgwi, Fasica Birhane Gebrekidan, Joan Kinyua, Sonia Kgomo, have shown that resisting discrimination and exploitation at work comes at a cost. Data training firms have responded by firing such workers who have voiced their anger against inhumane working conditions for them. But this cost African workers are willing to pay to fight for their survival and future. Data workers shared experiences have become a source for solidarity. For Atzeni (2009), solidarity is inbuilt into the labour process with a feeling of commitment towards fellow workers. As Richard noted:

> *'We formed a close circle of people at work. We would meet regularly in person and share what we have experienced earlier. More like an anonymous self-help group to comfort ourselves. People would come in and tell everyone else in the group what is affecting them. Sharing these experiences helped us feel lighter and realising that there are other people who are feeling the same. I think this was like a therapy to us. It was entirely the consequence of the failure of the management to address our issues. It was their obligation, and they failed us.'*

Mophat also noted that working in an open plan office allowed each one of us to grow closer to each other and '*essentially develop shared solidarity'.* The management of Sama was aware of both their activities at workplace and public profile due to local interviews in the media. Senior executives told them they are bothering too much and creating problems at workplace. Both were threatened with their jobs that their *'contract can end anytime.*'

Mophat and Richard, along with a few other colleagues started developing the idea of a group that would advocate for the rights of workers in the technology space in Africa, particularly data work. This would later become Techworker Community Africa (TCA). While they did not have clear plans about this group initially, they both left Sama in 2022, and the TCA got registered in January 2024. Currently, TCA is non-profit organisation working to increase awareness about the toll of data work on young Africans who are joining the labour markets and coordinating a multi-party dialogue to develop a framework for advancing labour rights in the industry. It is important to remember here that organising and mobilising workers is not an end itself, but it is an ongoing process to build sustained collective power.

**Conclusions**

Exploitation is everyday reality for workers around the world, irrespective of the sectors or industry they work in. For centuries, hard manual labour found in the shop, factory floor or a farm was considered by many to be physically demanding on workers bodies. With the emergence of new digital economy services sector activities, physical effort required to perform certain tasks got reduced to some extent. But the application of cognitive skills at work increased. As Jamie Woodcock (2016) argues that factories exploited bodies of workers and call centres exploited the minds. This paper shows that the exploitation of workers' bodies and minds go hand in hand in the production networks of AI.

In the discussion above, the paper has shown how data work is taking physical and mental toll on young Kenyans. Not only data workers exist in economically insecure employment relations (e.g. low pay and a lack of social protection), but they also experience severe health impacts including exhaustion, insomnia, depression, PTSD, vertigo, etc.

This raises serious ethical and political questions about the ways in which AI is made globally, including in some of the economically marginalised locations such as Africa. The AI tools cannot be considered a force for good, if the very foundation of these systems is built on the exploitation and subjugation of people who make them possible. In the African context, with centuries' long history of European colonialism that exploited its natural resources and labour, there is a worry that AI technologies might be recreating some of those dynamics. This paper has pointed towards this new trend via the exploitation of labour power involved in data work in Kenya.

However, exploitation and injustices also breed resistance (Kelly, 1998). Despite tech firms trying to hide their production practices to control labour in their value chains, data workers have become actively involved in pushing back against these firms. Alongside TCA, new worker led associations have emerged both in Kenya and elsewhere, including Data Labellers Association, Africa Tech Workers Movement, African Tech Worker Rising, to name a few. Future research should pay closer attention to the ways in which data workers, not just in Africa but in other parts of the global South are developing new ways to resist AI. The process of developing worker power is a never-ending struggle. To paraphrase Marx, *workers of AI, unite.*

**Acknowledgements:** We owe a great deal of gratitude to the data workers who gave us their precious time and share their stories with us. Also, Stephanie Diepveen, Sharath Srinivasan, George Karekwaivanane, and participants of the AI in Eastern Africa Workshop, Nairobi and European Conference of African Studies, Prague, and Digital Sovereignty Workshop, Stellenbosch (all 2025) provided useful feedback on earlier versions of this study, which greatly improved the paper.

**Funding statement:** The author would like to thank the UKRI (MR/Y017706/1) for funding authors' time and the Royal Society of Edinburgh (No. 3772) for funding the fieldwork.

**Disclosure statement:** The authors report there are no competing interests to declare.

**Figures:**

**1:** *Age of Kenyan Data Workers*

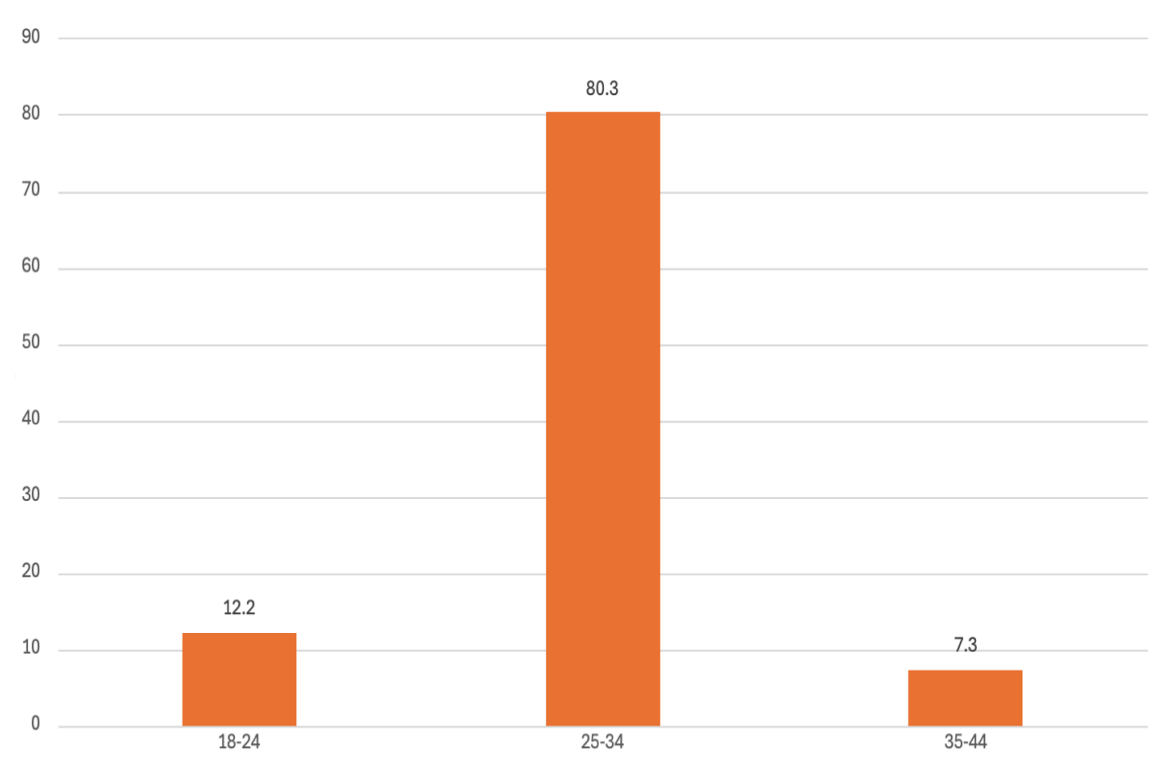

**2**: *Types of Data Work Tasks in Kenya.*

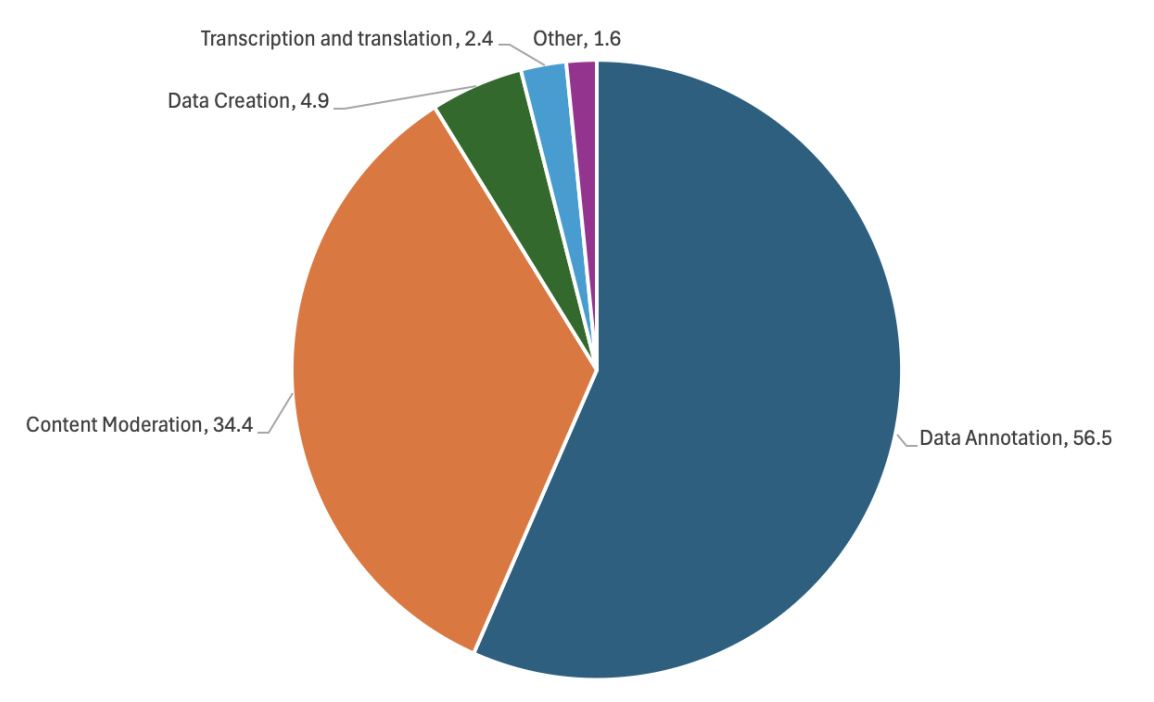

**3**: *Data workers' experiences at work*

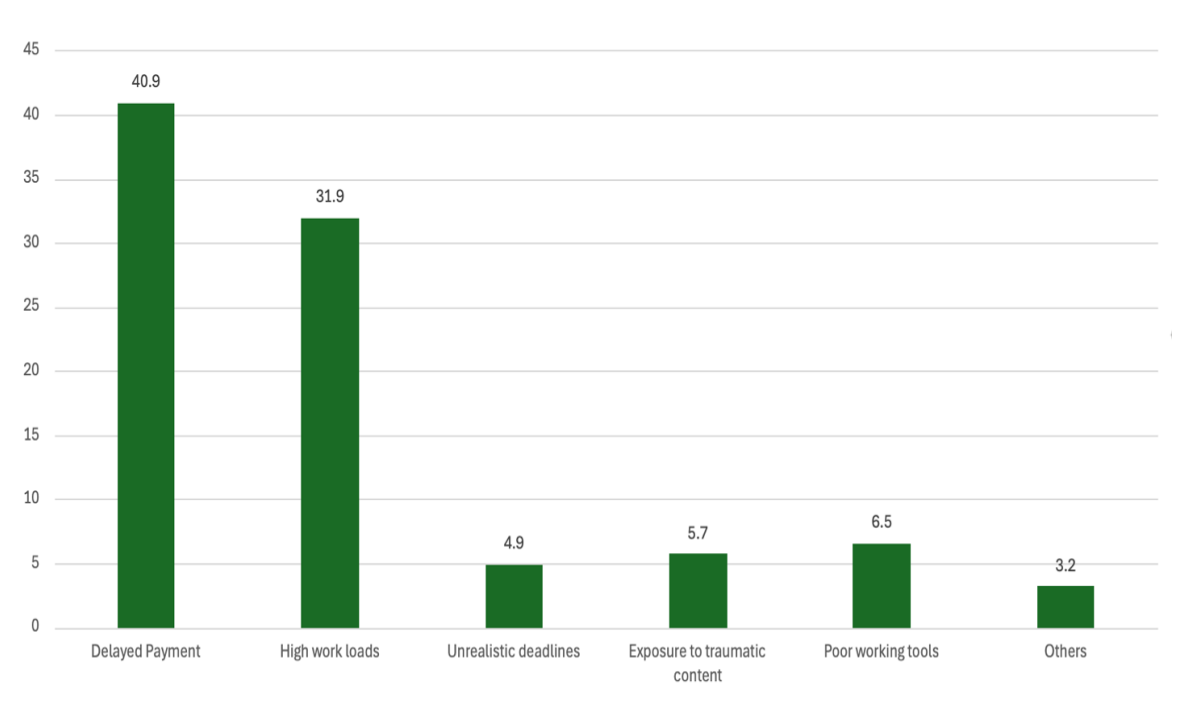

**4:** *Physical and Mental Health Impacts in Data Work.*

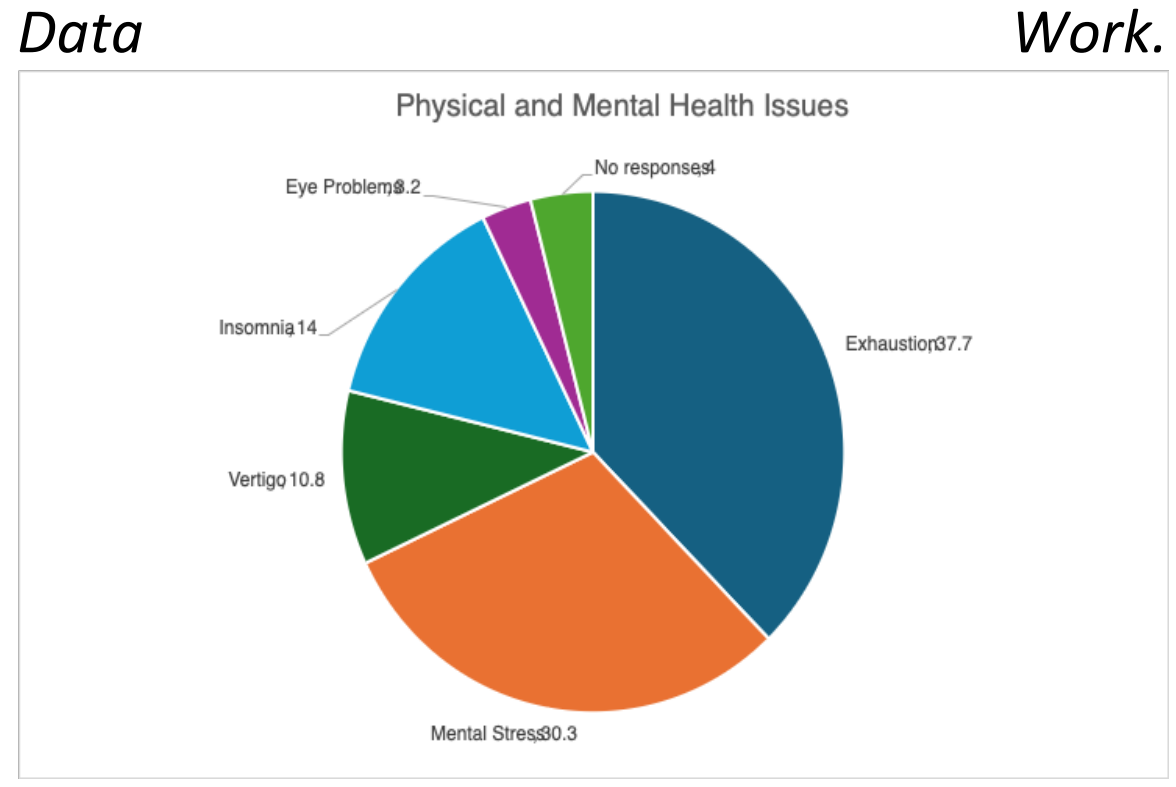